\documentclass[aps,prd,twocolumn,groupedaddress,showpacs]{revtex4}
\usepackage{amsmath, amssymb, bm}
\usepackage{amsfonts}
\usepackage{latexsym}
\usepackage{mathptmx}
\usepackage[mathscr]{eucal}
\usepackage{mathrsfs}
\usepackage[bf,small]{caption2}
\usepackage{float}
\usepackage{graphicx}
\DeclareMathSizes{10}{8}{6}{5}

\allowdisplaybreaks

\begin{document}


\title{Generalized Painlev\'{e}-Gullstrand descriptions of Kerr-Newman black holes}



\author{Huei-Chen Lin}
\email[l28981018@mail.ncku.edu.tw]{}
\affiliation{Department of Physics, National Cheng Kung University, Tainan 701, Taiwan}
\author{Chopin Soo}
\email[cpsoo@mail.ncku.edu.tw]{}
\affiliation{Department of Physics, National Cheng Kung University, Tainan 701, Taiwan}



\begin{abstract}
Generalized Painlev\'{e}-Gullstrand metrics are explicitly constructed for the Kerr-Newman family of charged rotating black holes.
These descriptions are free of all coordinate singularities; moreover, unlike the Doran and other proposed metrics, an extra tunable function is introduced
to ensure all variables in the metrics remain real for all values of the mass $M$, charge $Q$, angular momentum $aM$, and cosmological constant $\Lambda > -\frac{3}{a^2}$.
To describe fermions in Kerr-Newman spacetimes, the stronger requirement of non-singular vierbein one-forms at the horizon(s) is imposed and
coordinate singularities are eliminated by local Lorentz boosts. Other known vierbein fields of Kerr-Newman black holes are analysed and discussed; and it is revealed that
some of these descriptions are actually not related by physical Lorentz transformations to the original Kerr-Newman expression in Boyer-Lindquist coordinates - which is the reason
complex components appear (for certain ranges of the radial coordinate) in these metrics. As an application of our constructions the correct effective Hawking temperature for Kerr black holes
is derived with the method of Parikh and Wilczek.
\end{abstract}

\maketitle

\section{Overview}

As descriptions of charged rotating black holes, Kerr-Newman solutions \cite{Kerr1963,Newman1965} are important objects of study in General Relativity.
In Boyer-Lindquist coordinates \cite{Carter1968,Carter1973,Enderlein1997}, these metrics exhibit coordinate singularities at the horizon(s).
The Doran form of the solution \cite{Doran2000} can be considered to be the extension of the Painlev\'{e}-Gullstrand(PG) \cite{Gullstrand1922,Painleve1921} description
of a black hole from spherically symmetric to stationary axisymmetric spacetime. These descriptions have the advantage of being
free of coordinate singularities at the horizon(s). The Kerr solution, which was discovered much later, is not a
straightforward generalization of the Schwarzschild solution.  The Doran metric is likewise comparatively
recent; nevertheless, as ``regular" descriptions of rotating black holes,  the Doran and other proposed metrics have found their uses in black hole investigations.
Constant-time Doran slicings of the ergosurface in non-extremal black holes have been demonstrated to be free of conical singularities at the
poles \cite{Jacobson_Soong2009}. Calculations of Hawking radiation \cite{Parikh_Wilczek2000,Jiang2006}, and also neutrino asymmetry due to the interaction of fermions and rotating
black holes \cite{Mukhopadhyay2007} have also made explicit use of the Doran metric. Other authors have proposed to utilize the Doran metric to
extend or generalize spherically symmetric results to the context of rotating black holes \cite{Huhtala2002,Lasenby2005,Maluf2006}.

It is possible to eliminate the coordinate singularities at the horizon(s) by a choice of different coordinates; but if the final form is regular, then the coordinate transformation from the original singular form must be singular at the
horizon(s). In this work we achieve the stronger requirement of non-singular vierbein one-forms at the horizon(s). This not only guarantees the metric to be regular, but is in fact also a physical requirement in the presentation, for instance in Weyl or Dirac equations, of fermions in curved spacetimes. From this perspective, the coordinate singularities at the horizon(s) can be eliminated by Lorentz transformations of the vierbein one-forms.
Since rotations do not change the 3-geometry of constant-time slices, Lorentz boost(s) (with infinite rapidity at the horizon(s)) can be effective means to eliminate unphysical coordinate singularities.
Moreover deformation parameters of the 3-geometries of constant-time slices can also be introduced through these local boosts.

Recently it has been demonstrated \cite{Lin_Soo2009} that there is an obstruction to the implementation of flat PG slicings for
spherically symmetric spacetimes; and insistence on spatial flatness can lead to unphysical PG metrics with complex variables in which
the corresponding vierbein fields are not related to those of the standard spherically symmetric metric by physical Lorentz boosts.
Since the Doran form contains spherically symmetric PG solutions as special cases (the Reissner-Nordstr\"{o}m solution is an explicit example), it will be afflicted with similar problems \cite{Lin_Soo2009}.
In calculations of black hole evaporation using the Parikh-Wilczek method \cite{Parikh_Wilczek2000}, insistence on spatially flat PG coordinates can lead to spurious contributions which are ambiguous and problematic, both to the computation of the tunneling rate and to the universality of the results. In a more general context, the appearance of complex metric components causes unnecessary complications, and gives rise to difficulties
and ambiguities in the physical interpretations. As we shall demonstrate, other proposals \cite{Natario2008,Zhang_Zhao2005} of ``regular" Kerr-Newman black holes also suffer from similar problems.
In fact these descriptions and the Doran vierbein are actually not always related by physical Lorentz transformations to the original Kerr-Newman expression in Boyer-Lindquist coordinates - which is the reason
complex components appear (for certain ranges of the radial coordinate) in these metrics.

These troubles can be avoided altogether by using a less restrictive form of constant-time slicing which generalizes the Doran metric. We demonstrate how this goal can be realized, and construct a whole class of generalized (with adjustable function $f(r)$) real PG metrics for Kerr-Newman black holes which are completely free of coordinate singularities. In contradistinction, in our metrics no complex components arise for all values of $r$.
Although it is possible to introduce many parameters through the freedom of local Lorentz transformations which relate vierbein one-forms of the same metric, our generalized PG description is ``optimal'' in that only one additional function, $f(r)$, is needed, and introduced, to both reveal and avoid all the troubles. Different choices of $f$ result in different constant-$t_P$ slices of 3-geometry.
Our construction also recovers other known descriptions, and the method is used to clarify the relation between these metrics. Although the discussion of Hawking radiation is not the main theme of this work, we apply our constructions to the computation of Hawking radiation following the work of Parikh and Wilczek \cite{Parikh_Wilczek2000}. The correct result is obtained for both the Eddington-Finkelstein and our generalized PG metrics.

\section{Generalized Painlev\'{e}-Gullstrand metrics for Kerr-Newman spacetimes}

In this work we also include the contribution of non-trivial cosmological constant $\Lambda$ in Kerr-Newman solutions of rotating black holes with angular momentum $aM$ and charge $Q$.
Expressed in Boyer-Lindquist coordinates \cite{Carter1968,Carter1973,Enderlein1997}, the metric for the Kerr-Newman black hole is \footnote{Throughout this work geometric units $G=c=1$ and $(- + + +)$ spacetime signature are adopted.}
\begin{align}
&ds^2=-\frac{\Delta}{\Xi^2\rho^2}(dt-a\sin^2\theta
d\phi)^2+\frac{\rho^2}{\Delta}dr^2+\frac{\rho^2}{\Xi_\theta}d\theta^2\cr
&\quad\quad+\frac{\Xi_\theta\sin^2\theta}{\Xi^2\rho^2}(R^2d\phi-adt)^2,\cr
\;\; &R^2 :=r^2+a^2,
\;\rho^2:=r^2+a^2\cos^2\theta, \;\Xi:=1+\frac{1}{3}\Lambda a^2,\cr &\Xi_\theta:=1+\frac{1}{3}\Lambda
a^2\cos^2\theta,\;\Delta:=R^2\big(1-\frac{1}{3}\Lambda
r^2\big)-2Mr+Q^2.
\end{align}\\[5pt]
In this form, the metric suffers from coordinate singularities at the horizon(s) where $\Delta =0$. On the other hand the singularity at $\rho =0$ is physical and the curvature diverges there.
The metric is also problematic whenever $\Xi_\theta =0$ is allowed by the parameters involved, but this does not arise (even for negative cosmological constant) provided
$\Lambda>-\frac{3}{a^2}\;\;(a\neq 0)$.

For Schwarzschild-(anti)de Sitter black holes, the explicit Lorentz boost(s) between the singular standard spherically symmetric form and the regular PG-type metric have been discussed in Ref.\cite{Lin_Soo2009} .
Our aim here is to obtain, for the more intricate case of Kerr-Newman black holes, a class of vierbein one-forms (hence metrics)
which are real and regular everywhere, except at the physical singularity $\rho=0$.
To wit, we seek generalized Painlev\'{e}-Gullstrand descriptions of Kerr-Newman solutions by boosting the original singular vierbein by
\begin{equation}e_{trial}=L_1\Big(\frac{\sqrt{f^2-\Delta}}{f}\Big)\cdot e_{BL}.\end{equation}
The simplified notation above denote $({e^A_\mu})_{trial}dx^\mu = [L_1(\frac{\sqrt{f^2-\Delta}}{f}\Big)]^A\,_B ({e^B_\nu})_{BL}dx^\nu$ with the Boyer-Lindquist vierbien one-forms of Eq. (1) being
\begin{eqnarray}
\{(e^{A=0,1,2,3}_\mu)_{BL}dx^\mu\}&=
\{ \frac{\sqrt{\Delta}}{\Xi\rho}(dt-a\sin^2\theta d\phi), \frac{\rho}{\sqrt\Delta}dr,
\frac{\rho}{\sqrt {\Xi_\theta}}d\theta,\cr
&\frac{\sqrt{\Xi_\theta}\sin\theta}{\Xi\rho}(R^2d\phi-adt)\};
\end{eqnarray}
and $L_1(\beta)$ represents a Lorentz boost  (Lorentz indices are denoted by uppercase Latin letters) in the first (A=1) direction with rapidity $\xi = \tanh^{-1}\beta$.
Consequently,
\begin{align}
\{e^{A=0,1,2,3}_{trial}\}=&\{\frac{f}{\Xi\rho}(dt-a\sin^2\theta d\phi+\frac{\rho^2\Xi\sqrt{f^2-\Delta}}{\Delta f}dr),\cr & \frac{\sqrt{f^2-\Delta}}{\Xi\rho}(dt-a\sin^2\theta d\phi)+\frac{f\rho}{\Delta}dr, \frac{\rho}{\sqrt{\Xi_\theta}}d\theta,\cr & \frac{\sqrt{\Xi_\theta}\sin\theta}{\Xi\rho}(R^2d\phi-adt)\}\cr
=&\{\frac{f}{\Xi\rho}(dt_P-a\sin^2\theta d\phi_P), \frac{\sqrt{f^2-\Delta}}{\Xi\rho}(dt_P-a\sin^2\theta d\phi_P)\cr &+\frac{\rho}{f}dr, \frac{\rho}{\sqrt{\Xi_\theta}}d\theta, \frac{\sqrt{\Xi_\theta}\sin\theta}{\Xi\rho}(R^2d\phi_P-adt_P)\};
\end{align}
\noindent\\
wherein the PG time and azimuthal coordinates are defined as
\begin{equation}
dt_P:=dt+\frac{R^2\Xi\sqrt{f^2-\Delta}}{\Delta f}dr,
d\phi_P:=d\phi+\frac{a\Xi\sqrt{f^2-\Delta}}{\Delta f}dr.
\end{equation}
\noindent\\
Provided $f$ depends only on $r$, these are exact differentials.

To further eliminate $dt_P$ in $e^3_{trial}$ we can apply another
Lorentz boost $L_3\Big(\frac{a\sqrt{\Xi_\theta}\sin\theta}{f}\Big)$
resulting in
\begin{align}e_{GPG}=&L_3\Big(\frac{a\sqrt{\Xi_\theta}\sin\theta}{f}\Big)\cdot L_1\Big(\frac{\sqrt{f^2-\Delta}}{f}\Big)\cdot e_{BL}\cr
=&\{\frac{\sqrt{f^2-a^2\Xi_\theta\sin^2\theta}}{\Xi\rho}dt_P+\frac{a\sin^2\theta[\Xi_\theta
R^2-f^2]}{\Xi\rho\sqrt {f^2-a^2\Xi_\theta\sin^2\theta}}d\phi_P,\cr
&\frac{\sqrt{f^2-\Delta}}{\Xi\rho}(dt_P-a\sin^2\theta d\phi_P)
+\frac{\rho}{f}dr, \frac{\rho}{\sqrt{\Xi_\theta}}d\theta,\cr
&\frac{\rho\sqrt{\Xi_\theta}\sin\theta
f}{\Xi\sqrt{f^2-a^2\Xi_\theta\sin^2\theta}}d\phi_P\}.
\end{align}
\noindent\\
Note that the vierbein is now regular at horizon(s) $\Delta=0$, {\it
real}, and the metric Lorentzian $(-,+,+,+)$ in signature provided the
adjustable parameter $f(r)$ satisfies, at each value of $r$, the
criterion
\begin{equation}
f^2(r) >\max\{\Delta(r), a^2\Xi\}.
\end{equation}
This criterion is also precisely the condition which guarantees that
the Lorentz boosts in Eq.(6) have physical real rapidity parameters; and
we have thus obtained a class of generalized (with
adjustable parameter $f$) regular PG metrics for Kerr-Newman
solutions \footnote{The second Lorentz boost $L_3$, performed to eliminate $dt_P$ from $e^3_{GPG}$ and also to allow comparison with various
existing descriptions, is actually not needed to obtain a regular metric; in which case, following the steps leading to Eq.(4), criterion (7) can be relaxed to $f^2(r) > \Delta(r) \, \forall\, r$.}
. Different choices of $f$ result in different
constant-$t_P$ slices of 3-geometry. Moreover, as discussed, the
Lorentz boost becomes infinite (with $\beta = \tanh\xi =1$)
precisely at the horizon(s) (this is also the motivation for our
particular parametrization of $\beta$ in Eq.(2)).

An explicit, but by no means the only, choice for $f$ which
satisfies (7) is\begin{equation}
 f=\left\{\begin{array}{lll}\sqrt{R^2+Q^2 +\frac{\Lambda a^4}{3}},\;\;\;\; & \Lambda \geq 0\\
\sqrt{R^2(1-\frac{\Lambda}{3}r^2)+Q^2},\;\;\;\; & \Lambda < 0.\end{array} \right.\end{equation}
The explicit vierbein for positive cosmological constant is then \begin{align}
e_{GPG} =
&\{\frac{\sqrt{\rho^2+Q^2+\frac{\Lambda}{3}a^4(1-\sin^2\theta\cos^2\theta)}}{\Xi\rho}dt_P\cr
&+\frac{a\sin^2\theta[\Xi_\theta R^2-(R^2+Q^2)]}{\Xi\rho\sqrt{\rho^2+Q^2+\frac{\Lambda}{3}a^4(1-\sin^2\theta\cos^2\theta)}}d\phi_P,\cr
& \frac{\sqrt{2Mr+\frac{\Lambda}{3}(a^4+ R^2r^2)}}{\Xi\rho}(dt_P-a\sin^2\theta d\phi_P)\cr
&+\frac{\rho}{\sqrt{R^2+Q^2 +\frac{\Lambda a^4}{3}}}dr, \frac{\rho}{\sqrt{\Xi_\theta}}d\theta, \cr
&\frac{\rho\sqrt{\Xi_\theta}\sin\theta \sqrt{R^2+Q^2+\frac{\Lambda a^4}{3}}}
{\Xi\sqrt{\rho^2+Q^2+\frac{\Lambda}{3}a^4(1-\sin^2\theta\cos^2\theta)}}d\phi_P\}.
\end{align}

Our construction recovers several known solutions: When $f=R,
\Lambda=0$, we recover the Doran metric \cite{Doran2000} which is
compatible with the vierbein
\begin{align}
e_{Do}=&\{dt_P, \frac{\rho}{R}dr+\frac{\sqrt{R^2-\Delta}}{\rho}(dt_P-a\sin^2\theta d\phi_P), \rho d\theta,\cr & R\sin\theta d\phi_P\}\\
e_{Do}=&L_3\Big(\frac{a\sin\theta}{R}\Big)\cdot L_1\Big(\frac{\sqrt{R^2-\Delta}}{R}\Big)\cdot e_{BL}.
\end{align}
\noindent\\
However, for the Doran metric unphysical complex $t_P$ and $\phi_P$ arise for values of $r$ such that  $R^2-\Delta=2Mr-Q^2<0$ when $Q\neq 0$.
Also, in agreement with our earlier general understanding, the Doran form is then related to the Kerr-Newman solution in Boyer-Lindquist coordinates by an unphysical Lorentz boost
(as can be seen from complex rapidity parameter in the transformation (11) above). This deficiency can be overcome by our more general choice (or the explicit form in (8)) which satisfies criterion (7).
Without the advantage of a tunable function it is both hard to reveal the problem and also to guess an exact explicit form of $f(r)$ which renders the metric finite and real for all values of $r > 0$.

For the case of non-rotating Schwarzschild-(anti)de Sitter black
holes ($Q=a=0$), our generalized PG metric reduces to the form found
previously in Ref.\cite{Lin_Soo2009}, \begin{equation*}
ds^2=-\Big(\frac{f}{r}\Big)^2dt_P^2+\Bigg(\frac{r}{f}dr+dt_P\sqrt{\Big(\frac{f}{r}\Big)^2-1 + \frac{2GM}{r} + \frac{\Lambda R^2}{3}}\Bigg)^2+r^2d\Omega^2.
\end{equation*}

We can also recover the Eddington-Finkelstein description of
Kerr-Newman solutions by choosing a different set of time and
azimuthal coordinates, $dt_{EF}:= dt + \frac{\Xi R^2}{\Delta}dr$,
$d\phi_{EF}:= d\phi + \frac{\Xi a}{\Delta}dr$, after the first step
of Eq. (4). This results in \begin{align}
e_{EF}=&\{\frac{f}{\Xi\rho}(dt_{EF}-a\sin^2\theta
d\phi_{EF})+\frac{\rho(-f+\sqrt{f^2-\Delta})}{\Delta}dr,\cr &
\frac{\sqrt{f^2-\Delta}}{\Xi\rho}(dt_{EF}-a\sin^2\theta
d\phi_{EF})+\frac{\rho(f-\sqrt{f^2-\Delta})}{\Delta}dr,\cr &
\frac{\rho}{\sqrt{\Xi_\theta}}d\theta,
\frac{\sqrt{\Xi_\theta}\sin\theta}{\Xi\rho}(R^2d\phi_{EF}-adt_{EF})\}.
\end{align}
The vierbein is regular at the horizon(s) provided
$f-\sqrt{f^2-\Delta}=\alpha\frac{\Delta}{R}$ i.e.
$f=\frac{\alpha\Delta}{2R}+\frac{R}{2\alpha}$.
The criterion $f^2-\Delta>0$ needed to ensure reality of the
coordinates and the Lorentz boost can be attained by setting
\begin{equation}
 \alpha=\left\{\begin{array}{lll}\frac{R}{2\sqrt{R^2+Q^2}},\;\;\;\; & \Lambda\geq 0\\
\frac{R}{2\sqrt{R^2(1-\frac{\Lambda}{3}r^2)+Q^2}},\;\;\;\; & \Lambda < 0.\end{array} \right.\end{equation}
For positive cosmological constant,  we thus obtain the regular vierbein through $e_{EF}= L_1\left(\frac{4(R^2+Q^2)-\Delta}{4(R^2+Q^2)+\Delta}\right)\cdot e_{BL}$, giving\begin{align}
e_{EF}=&\{\frac{4(R^2+Q^2) + \Delta}{4\sqrt{R^2+Q^2}\Xi\rho}(dt_{EF}-a\sin^2\theta d\phi_{EF})-\frac{\rho}{2\sqrt{R^2+Q^2}}dr, \cr
&\frac{4(R^2+Q^2)-\Delta}{4\sqrt{R^2+Q^2}\Xi\rho}(dt_{EF}-a\sin^2\theta d\phi_{EF})
+\frac{\rho}{2\sqrt{R^2+Q^2}}dr, \cr &\frac{\rho}{\sqrt{\Xi_\theta}}d\theta, \frac{\sqrt{\Xi_\theta}\sin\theta}{\Xi\rho}(R^2d\phi_{EF}-adt_{EF})\}.
\end{align}
This yields Kerr black holes in the form of advanced Eddington-Finkelstein coordinates (for $\Lambda=0$ and $Q=0$) as \begin{align}
ds^2&=\eta_{AB}e^A_{EF}e^B_{EF}\cr&=-(1-\frac{2Mr}{\rho^2})dt_{EF}^2+2dt_{EF}dr\cr&-\frac{4Mra\sin^2\theta dt_{EF}d\phi_{EF}}{\rho^2}-2a\sin^2\theta drd\phi_{EF}\cr
&+\rho^2d\theta^2 +(R^2\sin^2\theta +\frac{2Mra^2\sin^4\theta}{\rho^2})d\phi_{EF}^2,
\end{align}
which is free of coordinate singularities and always real.

\section{Hawking radiation}

As an application of our constructions, we consider Hawking radiation for Kerr black holes with the tunneling process of Parikh and Wilczek \cite{Parikh_Wilczek2000}.
In such computations, regularity of the metric at the horizon(s) is essential; singular metrics can lead to factor-of-two discrepancies and other complications \cite{Luciano_europhys}.
We shall demonstrate that both the regular advanced Eddington-Finkelstein metric and our generalized PG metrics yield the correct results.
To wit, Hawking radiation is treated as tunneling across the (outer) horizon from ${\bf r}_{in}$ to ${\bf r}_{out}$ of massless emissions carrying energy and angular momentum.
The black hole mass parameter $M$ shrinks by $\omega$ and the angular momentum changes by $a\omega$ resulting in  $TdS_{BH} = dM - \Omega dJ =-\omega(1-\Omega a)$,
for the first law of black holes \cite{Hawking_1975_Commun.}. $\Omega$ is the angular velocity at the outer horizon $r_+ = M + \sqrt{M^2 + a^2}$.
The decay rate comes from the imaginary part of the associated particle action \cite{Parikh_Wilczek2000} (for simplicity we consider massless particles without spin) which is
\begin{eqnarray}
I&=&\int^{\bf{r}_{out}}_{\bf{r}_{in}} {\bf{p}\cdot \bf{dr}}=\int^{\bf{r}_{out}}_{\bf{r}_{in}}\Big(\int^{\bf{p}}_0 {\bf dp}\Big)\cdot{\bf dr}\cr
&=&\int^{r_{out}}_{r_{in}}
\Big(\int^{H_0+\omega(1-\Omega a)}_{H_0} \frac{dH}{{\dot{r}}^{\,i}}\Big)dr^i.
\end{eqnarray}\\
In the last step, Hamilton's equation, $\left.\frac{dH}{dp^{\;i}}\right|_{\bf r}={\dot{r}}^{\,i}$ , for the semiclassical process is invoked. Switching the order of integration, together with $dH =d\omega'(1-\Omega' a)$, yields
\begin{equation}
I=\int^{\omega}_{0}\int^{r_{out}}_{r_{in}} \frac{dr^{i}}{\dot{r}^{\;i}}(1-\Omega' a)d\omega', \qquad \Omega'=\frac{a}{r^2_{+}(\omega')+a^2},
\end{equation}\\
with $r_+(\omega') := (M-\omega') + \sqrt{(M-\omega')^2 + a^2}$ denoting the location the shifted horizon when $M$ decreases by $\omega'$.

With advanced Eddington-Finkelstein coordinates, the null geodesics corresponding to $\{\frac{dt_{EF}}{dp},\frac{dr}{dp},\frac{d\theta}{dp},\frac{d\phi_{EF}}{dp}\}=\{\frac{R^2}{\Delta},\frac{1}{2},0,\frac{a}{\Delta}\}$ yield $\dot{r}=(\frac{dr}{dp})/(\frac{dt_{EF}}{dp})=\frac{\Delta}{2R^2}$, and thus \begin{equation}
I=\int^{\omega}_{0}\int^{r_{out}}_{r_{in}} dr\frac{2R^2}{\Delta}(1-\Omega'a)d\omega'+I_{\phi_{EF}};
\end{equation}
 wherein $I_{\phi_{EF}}:=\int^{\omega}_{0}\int^{{\bf r}_{out}}_{{\bf r}_{in}} [{d\phi_{EF}}/{(\dot{\phi}_{EF})}](1-\Omega' a)d\omega'$ does not contribute to the imaginary part of $I$ as
 $\dot{\phi}_{EF}=\frac{a}{R^2}$ is finite at the outer horizon.
The integral $I$, with pole at $r_+$, is defined by deforming the contour to go through a clockwise
infinitesimal semicircle $r = r_{+} +\epsilon e^{i\theta}$ around the pole \cite{Parikh_Wilczek2000}. Its imaginary part is then \begin{eqnarray}
\Im I&=&\Im\left.\lim_{\epsilon\to 0} \int^{\omega}_{0}\int^{\pi}_{2\pi} \epsilon e^{i\theta}id\theta\Big[\frac{2(R^2+2r\epsilon e^{i\theta}+\cdots)}{\Delta+\epsilon e^{i\theta}\partial_r \Delta+\cdots}\Big]\right|_{r=r_{+}(\omega')}(1-\Omega'a)d\omega'\cr
&=&-\int^{\omega}_{0}\left.\frac{2\pi(r^2+a^2)}{r-r_{-}}\right|_{r=r_{+}(\omega')}(1-\Omega'a)d\omega'
\end{eqnarray}

For our generalized PG metrics (4), it can be verified that $\{\frac{dt_P}{dp},\frac{dr}{dp},\frac{d\theta}{dp},\frac{d\phi_P}{dp}\}=e^{\int^r g(r') dr'}\{R^2,\frac{f\Delta}{f+\sqrt{f^2-\Delta}},0,a\}$
correspond to null geodesics if $ g :=\frac{f(\partial_r \Delta)(f^2+f\sqrt{f^2-\Delta}-\frac{\Delta}{2})-(\partial_r f)\Delta^2}{f\Delta(\Delta-f^2-f\sqrt{f^2-\Delta})}.$ Now with  $\dot{r}=(\frac{dr}{dp})/(\frac{dt_P}{dp})=\frac{f\Delta}{R^2(f + \sqrt{f^2-\Delta})}$, and proceeding as before, we are lead to
\begin{equation}
I(\omega)=\int^{\omega}_{0}\int^{r_{out}}_{r_{in}} dr\frac{R^2(f+\sqrt{f^2-\Delta})}{f\Delta}(1-\Omega'a)d\omega' + I_{\phi_P}.
\end{equation}
 Provided $\sqrt{f^2-\Delta}$ remains real, the imaginary part of the action is \begin{eqnarray}
 \Im I&=&\Im\lim_{\epsilon\to 0}\int^{\omega}_{0}\int^{\pi}_{2\pi}d(i\theta)(1-\Omega'a)d\omega'\cdot\cr
 &&\epsilon e^{i\theta}\left.\Big[\frac{(R^2+\epsilon e^{i\theta}2\partial_r R)[f+\sqrt{f^2-\Delta}+\epsilon e^{i\theta}\partial_r(f+\sqrt{f^2-\Delta})]}{f\Delta+\epsilon e^{i\theta}(f\partial_r\Delta+\Delta\partial_r f)}\Big]\right|_{r=r_{+}(\omega')}\cr
 &=&-\int^{\omega}_{0}\left.\pi\Big[\frac{R^2(f+\sqrt{f^2-\Delta})}{f\partial_r \Delta}\Big]\right|_{r=r_{+}(\omega')}(1-\Omega'a)d\omega'\cr
 &=&-\int^{\omega}_{0}\left.\frac{2\pi(r^2+a^2)}{r-r_{-}}\right|_{r=r_{+}(\omega')}(1-\Omega'a)d\omega'.
\end{eqnarray}
Both Eqs. (19) and (21) give the {\it same} final result. The change of the Bekenstein-Hawking entropy from $\Delta S_{BH} =-\Delta S=2\Im I(\omega)$ yields, at the lowest order (higher order corrections in $\Im I(\omega)$ indicate departures from pure thermal physics), the effective temperature as \begin{equation*}T_{eff}=\left.\Big[\frac{dS_{BH}}{(-dH)}\Big]^{-1}\right|_{\omega =0}=\left.\Big[\frac{2d\Im I}{-(1-\Omega a)d\omega}\Big]^{-1}\right|_{\omega =0}=\frac{r_+-r_{-}}{4\pi(r_+^2+a^2)},\end{equation*}
which agrees with the Hawking temperature $T_{Hawking}= \frac{\kappa}{2\pi }=  \frac{r_+-r_{-}}{4\pi(r_+^2+a^2)}$.
Thus both the Eddington-Finkelstein and our generalized PG metrics lead to the same physical effective Hawking temperature.
It should be noted that for the generalized PG metrics, the result is, reassuringly, {\it independent} of the function $f$, provided the physical criterion $f^2-\Delta > 0$ is maintained.

\section{Discussion and further remarks}

The constant-$t_P$ hypersurfaces are conformally flat iff the
Cotton-York tensor vanishes \cite{York1971}. The explicit computation
of the tensor for the constant-$t_{P}$ 3-geometries exhibited here is rather involved
and we have not been able to find any choice of $f$ which results in conformally flat slicings.
There is analytic proof that there is no maximal, non-boosted, conformally
flat slices not only in the Kerr spacetime, but also in any
stationary spacetime with non-vanishing angular
momentum \cite{Valiente2004}. For the special case of spherically
symmetric metrics, PG metrics with flat slicing can violate
criterion (7). This can be remedied by appropriate choices of $f$
which however does not in general lead to spatial
flatness \cite{Lin_Soo2009}.

With vanishing cosmological constant, the Doran metric has
$e^0_{Do}\propto dt_P$ but, as discussed, unphysical complex
components appear for certain values of $r$ whenever $Q\neq 0$. In
the class of metrics we constructed with regular real vierbein for
all values of $r$,  the requirement $e^0_{GPG}\propto dt_P$ is
satisfied iff $R^2\Xi_\theta-f^2=0$. The latter cannot be
achieved for non-vanishing cosmological constant for our vierbeins since
$\frac{\partial f}{\partial \theta}$ is required to vanish;
the gauge condition ${e'}^0_{GPG}\propto dx^0$ can however be attained by a further Lorentz boost.

Other parametrizations of Kerr-Newman solutions have been suggested, and our general understanding can also be applied to analyze these prescriptions.
An alternative to the Doran metric, proposed by Natario \cite{Natario2008} for pure Kerr black holes, is
\begin{equation*}
e_{\mathrm{Nat}}=\{
dt_{P},\frac{\rho}{\sqrt\sigma}(dr-vdt_{P}),\rho
d\theta,(d\phi_{P}+\delta d\theta-\Omega_C
dt_{P}){\sqrt\sigma}\sin\theta\},\end{equation*}
 with $dt_{P}=dt-k dr$, $d\phi_{P}=d\phi-k\Omega_C dr -\delta d\theta$,
$\rho^2 v=-R\sqrt{R^{2}-\Delta}, \rho^2\sigma= R^4-a^2\Delta\sin^2\theta, k =\frac{\rho^{2}v}{\Delta}, $ and
$\delta=-\int^\infty_r k\frac{\partial\Omega_C}{\partial\theta}dr$.
It is related to the Kerr-Newman metric in Boyer-Lindquist coordinates by\[
e_{\mathrm{Nat}}=L_{1}\left(\frac{R}{\rho}\sqrt{\frac{R^{2}-\Delta}{\sigma}}\right)\cdot
L_{3}\left(\frac{a\sqrt{\Delta}\sin\theta}{R^{2}}\right)\cdot
e_{\mathrm{BL}}.\] However, without the benefit of our adjustable function $f$ in the boost to ensure criterion (7) is satisfied, the problem with the above metric is again the condition $R^{2}-\Delta>0$ for physical Lorentz
boosts and real metric variables is violated for some values of $r$. Another proposed metric \cite{Zhang_Zhao2005} also suffers from similar problems.

It is possible to discuss Kerr-Newman solutions within the context of general axisymmetric metrics expressed in Chandrasekhar form \cite{Chandrasekhar_Friedman1972,Chandrasekhar1992}. The latter is compatible with the vierbein
\begin{equation}
e_{\mathrm{axisym}}=\left\{ A_{s}dt,\quad B_{s}^{-1}dr,\quad\rho
d\theta,\quad C_{s}(d\phi-\Omega_C dt)\right\},
\label{eq:e_ax}\end{equation}
with Kerr-Newman parameters,
\[ \begin{array}{lll}
A_{s}=\frac{\sqrt{\Delta}\rho\Xi_{\theta}}{\Xi\sqrt{R^{4}\Xi_{\theta}^{2}-a^{2}\Delta\sin^{2}\theta}},  &B_{s}=\frac{\sqrt{\Delta}}{\rho}\\ & \\
C_{s}=\frac{\sqrt{R^{4}\Xi_{\theta}^{2}-a^{2}\Delta\sin^{2}\theta}}{\rho\Xi}\sin\theta,
&\Omega_C=\frac{a(R^{2}\Xi_{\theta}^{2}-\Delta)}{R^{4}\Xi_{\theta}^{2}-a^{2}\Delta\sin^{2}\theta}.
\end{array}\]
However, the Kerr-Newman solution in Chandrasekhar form has extra coordinate singularities at $\Sigma^2=R^4\Xi_\theta-a^2\Delta\sin^2\theta=0$.
The reason is again revealed by our general understanding between singular and regular forms of the veirbein. For Kerr-Newman metrics, the explicit relation between the Chandrasekhar and Boyer-Lindquist expressions
is again a Lorentz boost, $e_{axisym}=L_3\Big(\frac{a\sqrt\Delta\sin\theta}{R^2\sqrt{\Xi_\theta}}\Big)\cdot e_{BL},$
which is infinite at $\Sigma^2=0$. Thus the above Chandrasekhar form of Kerr-Newman black holes will have additional coordinate singularities at $\Sigma^2=0$ to contend with, in addition to the coordinate singularity of
 the Boyer-Lindquist expression (1) at the horizon(s) $\Delta =0$. In contradistinction, our generalized PG expressions of Kerr-Newman black holes constructed and displayed in this work are free of all coordinate singularities, and real, for all values of $r$.

\section*{Acknowledgments}
This work has been supported in part by the National Science Council
of Taiwan under Grant Nos. NSC98-2112-M-006-006-MY3 and 99-2811-M-006-015, and by the National Center
for Theoretical Sciences, Taiwan. Beneficial interactions with C. Y. Lin during the early phase of this work are also gratefully acknowledged.

\end{document}